 \documentclass[final,5p,times,twocolumn,authoryear]{elsarticle}

\usepackage{amssymb}
\usepackage{lipsum}
\usepackage{amsmath}
\journal{Physics Letters B}

\begin{document}

\begin{frontmatter}



\title{Study on the Contribution of Positronium and $\pi^{0}$ Mesons to Casimir Force}


\author[first]{Cong Li}
\affiliation[first]{organization={Zhejiang Ocean University},%Department and Organization
            addressline={2022132@zjou.edu.cn}, 
            city={Zhoushan},
            postcode={}, 
            state={Zhejiang},
            country={China}}

\begin{abstract}
There is a Casimir force between two metal plates. It is generally believed that the Casimir force is mediated by virtual photons in a vacuum, which correspond to the massless intermediate particles used in our theoretical calculations. Studies have shown that not only virtual photons in a vacuum, but also other virtual particles that have masses. The lightest chargeless virtual particles with mass are positronium (1 MeV) and $\pi^{0}$ mesons (135 MeV). This paper primarily focuses on studying the corrections to the Casimir force caused by positronium and  $\pi^{0}$ mesons. Especially when the distance between the two plates is on the order of $1/m_{positronium}$ , the contribution of positronium becomes significant, and on the order of $1/m_{\pi^0}$ , the contribution of the  $\pi^{0}$ meson becomes significant. We hope that the calculation results can reduce the error in the theoretical calculation of the Casimir force when the distance between the plates is large and provide significant corrections when the distance is small.
\end{abstract}




\end{frontmatter}

%\tableofcontents



\section{Introduction}
\label{introduction}

The Casimir effect\citep{Casimir:1948dh} describes the force that develops between boundaries in a finite space due to quantum fluctuations\citep{fosco2008functional} in the vacuum. There are two elements involved in its production: vacuum and limited space. In general, it is believed that quantum fluctuations are primarily caused by virtual photons\citep{Casimir:1948dh}. However, there are other wave-like virtual particles in the vacuum that can also contribute to quantum fluctuations. Unlike photons, these virtual particles can possess varying masses\citep{dirac1981principles}\citep{feynman2018theory}. According to the uncertainty principle, the lifetime of a virtual particle is inversely proportional to its energy. Therefore, the effect is expected to be observed only over a more limited spatial range due to its short life. This paper primarily studies the impact of mass on the Casimir force.

The Casimir effect of massive objects has been widely studied, as done in work\citep{1}, where the authors place two parallel plates in a massive bosonic background field to obtain the Casimir effect caused by the massive bosonic field. In work \citep{2}, the authors introduce the Casimir effect of massive particles in 1+1-dimensional and 1+2-dimensional spaces. This work focuses on studying the Casimir effect in a 4-dimensional vacuum background, which not only contains photons but also other particles, making it a complex background field. First, the carrier particles between the plates must satisfy certain conditions: they must not carry any charge, including electric charge, weak isospin, or even color charge, in order to ensure the neutrality of the two plates and the conservation of various charges. Since the contributions of massless particles has already been considered in previous works, this study mainly focuses on the corrections brought by massive particles. Additionally, this study mainly focuses on the distance between the plates from large to small. Among the particles with mass, we prioritize the consideration of the two lightest ones: positronium and $\pi^{0}$ mesons.

The structure of this paper is as follows. In the next section, we will compute the effect of mass on the Casimir force between two plates in 4-dimensional spacetime, and give numerical results. Then, we will further consider the correction of the Casimir force by the electronium and $\pi^{0}$\ mesons, and give numerical results. A brief summary will be given at the end.
\section{Influence of Mass on Casimir Force in 1+3-Dimensional Spacetime}
%%\label{}
Assuming the plane is parallel to the x-y plane and the distance is a, the wave function of the particles between planes is
\begin{equation}
\psi_n(x, y, z ; t)=e^{-i \omega_n t} e^{i k_x x+i k_y y} \sin \left(k_n z\right)
\end{equation}
where $k_n=\frac{n \pi}{a}(n=0,1,2 \ldots)$ and $\omega_n=\sqrt{m^2+k_x^2+k_y^2+k_n^2}$ Therefore, the vacuum energy of a virtual particle with mass between the plates is 
\begin{equation}
E_0=\langle H\rangle=\frac{1}{2} \int \frac{S d k_x d k_y}{(2 \pi)^2} \sum_{n=-\infty}^{\infty} \omega_n,
\end{equation}
where S is the area of the two plates. Obviously, the integral result above is infinite, but we can obtain a renormalized result by subtracting the vacuum energy in the absence of the plates. We have 
\begin{equation}
\begin{aligned}
E_0^{\mathrm{ren}}(a) & =\frac{1}{2} \int \frac{d k_x d k_y}{(2 \pi)^2}\left(\sum_{n=-\infty}^{\infty} \omega_{\mathbf{k}_{\perp}, n}-\int \frac{\mathrm{d} k_3}{2 \pi} \omega_{\mathbf{k}}\right) S \\
& =\frac{1}{2} \int \frac{d k_x d k_y}{(2 \pi)^2}\left(2 \sum_0^{\infty} \omega_{\mathbf{k}_{\perp}, n}-\omega_0-2 \int_0^{\infty} \frac{\mathrm{d} k_3}{2 \pi} \omega_{\mathbf{k}}\right) S  \\
& =\frac{1}{2} \int \frac{k_{\perp} \mathrm{d} k_{\perp}}{\pi}\left(\sum_0^{\infty} \omega_{\mathbf{k}_{\perp}, n}-\int_0^{\infty} \frac{\mathrm{d} k_3}{2 \pi} \omega_{\mathbf{k}}-\frac{\omega_0}{2}\right) S  \\
& =\frac{1}{2} \int \frac{k_{\perp} \mathrm{d} k_{\perp}}{\pi}\\
&\quad \times \left(-\frac{\omega_0}{2}-\frac{1}{2 \pi a} \int_\mu^{\infty} \frac{d y}{\exp (y)-1} \sqrt{y^2-\mu^2}-\frac{\omega_0}{2}\right) S \\
& =\int \frac{k_{\perp} \mathrm{d} k_{\perp}}{2\pi}\left(-\omega_0-\frac{1}{2 \pi a} \int_\mu^{\infty} \frac{d y}{\exp (y)-1} \sqrt{y^2-\mu^2}\right) S, 
\end{aligned}
\end{equation}
where $\omega_{\mathbf{k}}=\sqrt{m^2+k_x^2+k_y^2+k_z^2}$. The Abel-Plana formula is used in the fourth equation \citep{3}, where $y=2 \pi k_3$ and $\mu=2 \sqrt{m^2+k_{\perp}^2} a$. After taking the derivative of $a$, we obtain the Casimir force dependent on mass caused by a massive particle in 4-dimensional spacetime, given by 
\begin{equation}
\begin{aligned}
F_{\text {casimir }}(a, m)=&-\frac{1}{4 \pi^2 a^2} \int_0^{\infty} k_{\perp} \mathrm{d} k_{\perp} \int_\mu^{\infty} d y\\
&\times \left(\frac{\mu^2}{[\exp (y)-1] \sqrt{y^2-\mu^2}}+\frac{\sqrt{y^2-\mu^2}}{\exp (y)-1}\right)
\end{aligned}
\end{equation}
It is obvious that when we set m=0, the equation above will degenerate into the familiar Casimir force caused by massless photons in 4-dimensional spacetime \citep{4}. And then it is easy to verify,
\begin{equation}
F_{\text {casimir }}(a, 0)=\frac{\pi^2}{240 a^4}
\end{equation}
It shows that our mass-dependent Casimir force results are consistent with previous works when the mass is zero. Further analytical calculations seem difficult, but we can numerically obtain the dependence of the Casimir force on mass and plate distance.

Then we have considered four particles with masses of 1GeV, 2GeV, 3GeV and 4GeV respectively, and their dependence on the plate distance is shown in FIG. 1. In FIG. 1, it shows that among the four curves with different masses, the dependence of the force on the distance is monotonically decreasing, and the larger the mass, the faster the decrease.

\begin{figure}
	\centering 
	\includegraphics[width=0.45\textwidth, angle=0]{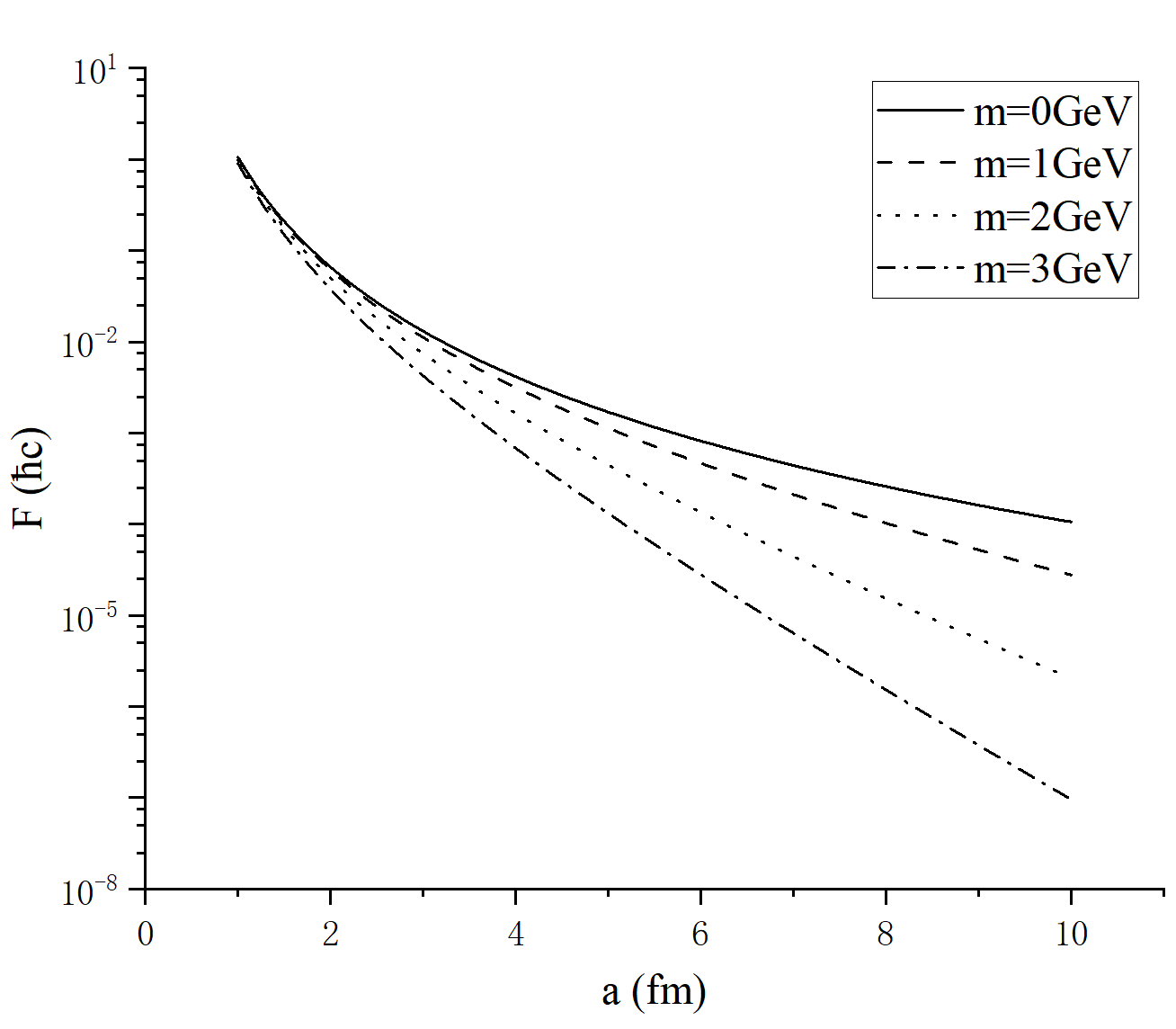}	
	\caption{The dependence of the Casimir force on distance for particles with masses of 0, 1, 2, and 3 GeV, respectively.} 
	\label{0-3}%
\end{figure}

\section{Correction of Casimir Force by positronium and $\pi^{0}$ Meson}

In the above part, it is shown that any massive particle can also cause Casimir force, including virtual particles such as virtual electrons in vacuum. Charged particles seem to be complex. We adopt two neutral bosons, positronium and $\pi^{0}$ Meson, to simplify further discussions.

We further consider the vacuum in field theory, which contains not only massless virtual photons, but also low-mass virtual positronium with a mass of 1 MeV. Therefore, the Casimir force between the two plates is influenced not only by photons but also by positronium. By considering these two types of particles and doing numerical calculations, we obtain Figure 2.
\begin{figure}
	\centering 
	\includegraphics[width=0.45\textwidth, angle=0]{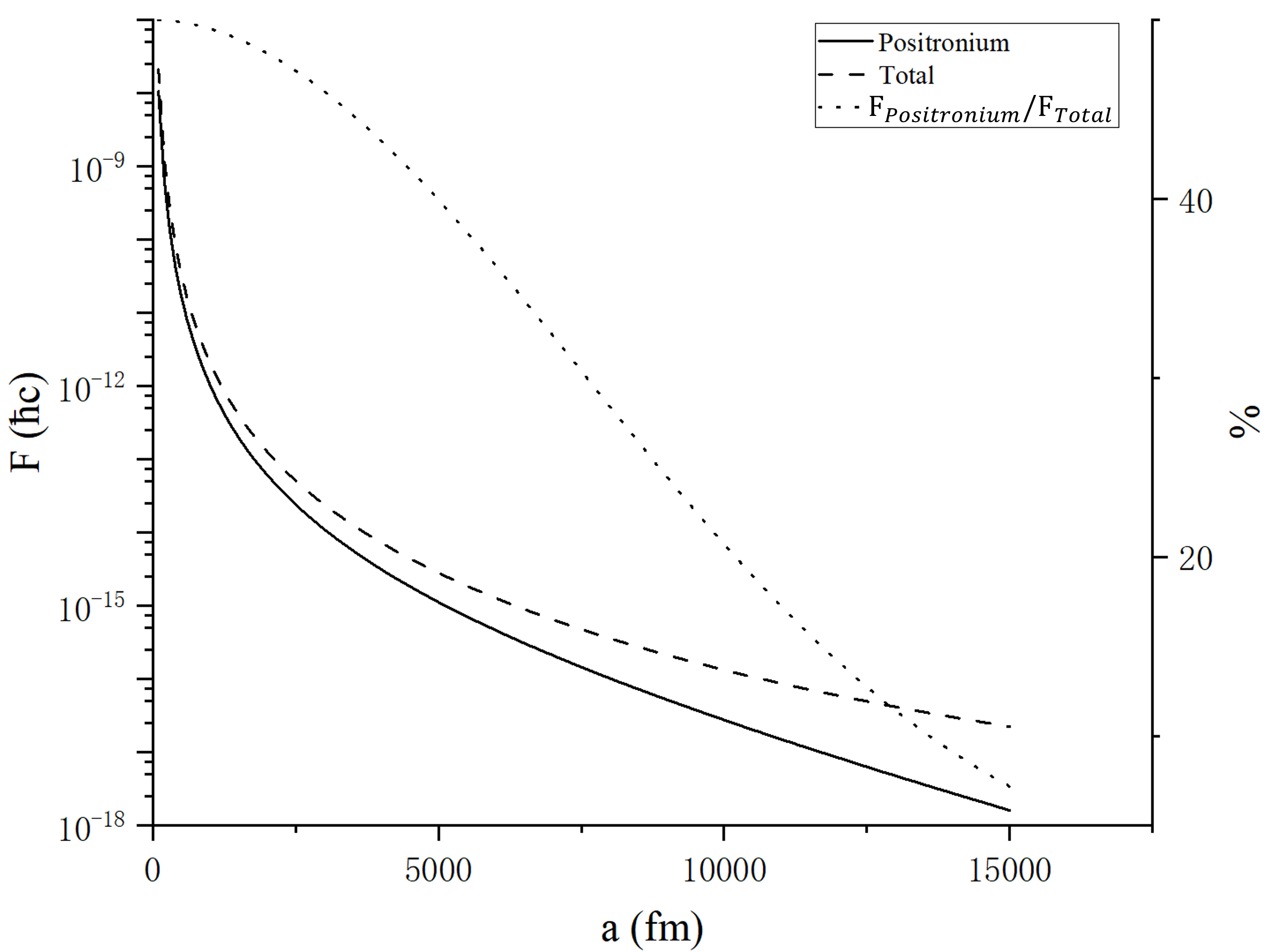}	
	\caption{The contributions of positronium and positronium-photon to the Casimir force, as well as the ratio between the two.} 
	\label{e-p}%
\end{figure}
The percentage curve of the positronium contribution shows that when the plate distance is very large ($>$10000fm), the contribution of the positronium can be ignored. At this point, considering only photons is sufficient and consistent with previous calculations. As the plate distance decreases to between 5000-10000fm, the contribution of the positronium becomes increasingly significant. When the plate distance is less than 5000fm, the contribution of the positronium becomes equally important as photons, and the Casimir force doubles in magnitude.

After taking into account the positronium, we can further introduce the contribution of $\pi^{0}$ mesons, and we can get the dependence of the Casimir force on the plate distance with $\pi^{0}$ mesons, as shown in Figure 3.
\begin{figure}
	\centering 
	\includegraphics[width=0.45\textwidth, angle=0]{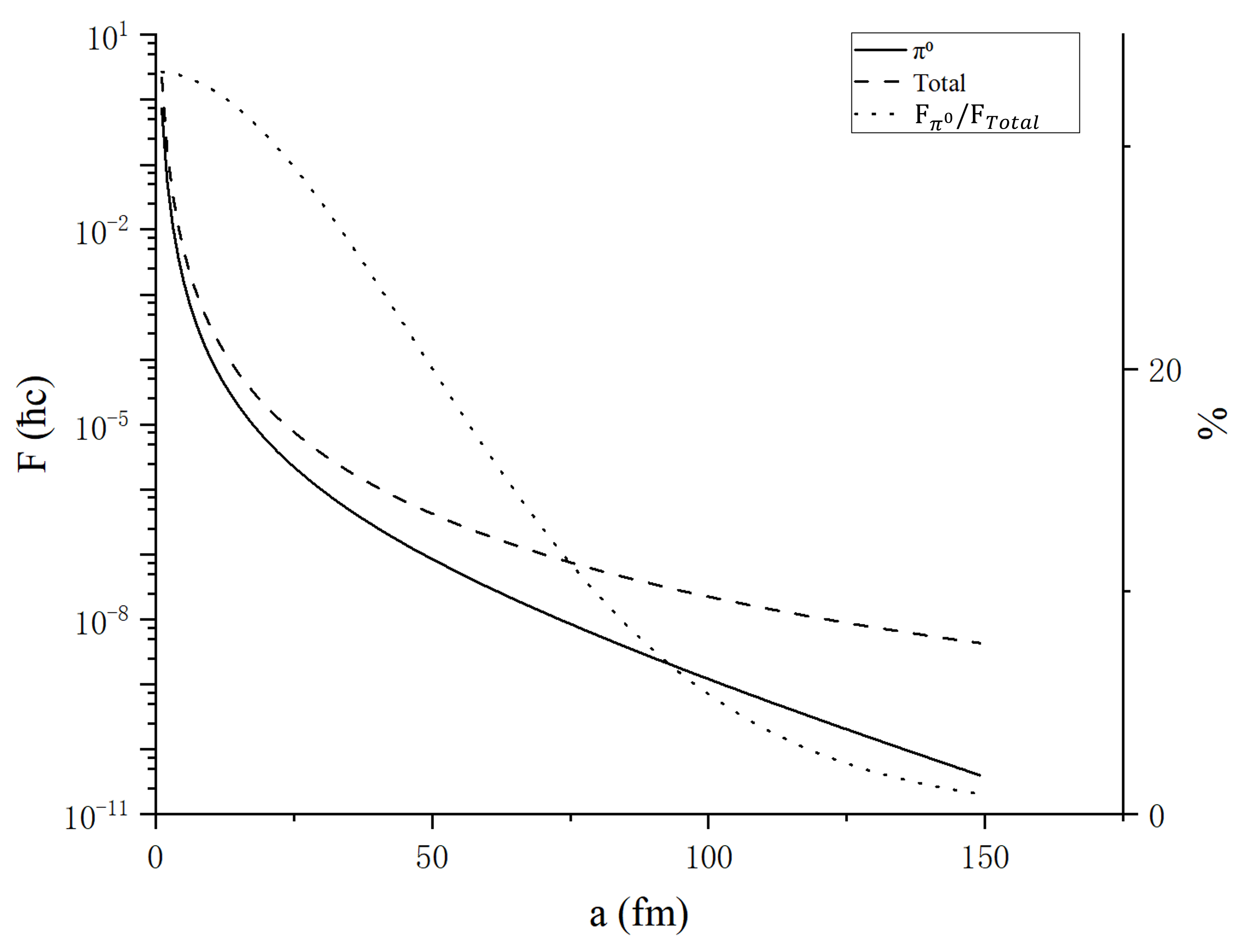}	
	\caption{The contributions of $\pi^{0}$ and $\pi^{0}$-positronium-photon to the Casimir force, as well as the ratio between the two.} 
	\label{pi0}%
\end{figure}
As shown in FIG. 3, after including the $\pi^{0}$ mesons, their contributions are very small and almost negligible when the distance is greater than 150 fm. As the distance decreases, their contribution becomes increasingly significant. When the distance is less than 10 fm, the Casimir force caused by $\pi^{0}$ mesons become as important as the forces caused by the other two particles, accounting for about 1/3.

\section{Summary and conclusions}
%%\label{}
In this work, we study the effect of mass on the Casimir force in four-dimensional spacetime. We find that the Casimir force has a negative correlation with the mass, that is, for a certain distance between the plates, the greater the mass, the smaller the force.

We further study that there are other virtual particles besides photons that can contribute to the Casimir force. Including the two lightest uncharged virtual particles, positronium and $\pi^{0}$ meson. As the plate distance decreases, the contribution of these particles increases significantly, eventually reaching a magnitude comparable to that of the photon. When the plate distance is between 1000 fm and 100 fm, the Casimir force is approximately twice as strong as that of the photon. At a distance near 10 fm, the Casimir force is roughly three times stronger than that of the photon. At even smaller distances, an increasing number of particles will contribute, leading to a multiplying effect on the force.

During the work, we also made an interesting discovery. Figure 2 and Figure 3 reveal that the plate distance at which the contribution of the two particles becomes significant is 10000 fm for positronium and 100 fm for the $\pi^{0}$ meson, and this distance is just equal to the reciprocal of the two masses. An interesting observation is the similarity to the condition of wave diffraction, where the wavelength is equivalent to the size of the obstacle.

\section*{Acknowledgements}
Thanks to Minglun-Mo, He-yao Zhang and Du-xin Zheng for helpful discussions.

\bibliographystyle{elsarticle-harv} 
\bibliography{example}



%% For example:

%% \bibitem[Aladro et al.(2015)]{Aladro15} Aladro, R., Martín, S., Riquelme, D., et al. 2015, \aas, 579, A101

%%\end{thebibliography}

\end{document}